\documentstyle[12pt]{article}
\input{epsfig.sty}
\newcounter{my}

\newcommand{\re}[1]{\ (\ref{#1})}
\newcommand{\nn}{\nonumber}
\newcommand{\ed}{\end{document}}
\newcommand{\be}{\begin{equation}}
\newcommand{\ee}{\end{equation}}

\newcommand{\ba}{\begin{eqnarray}}
\newcommand{\ea}{\end{eqnarray}}
\newcommand{\baz}{\begin{eqnarray*}}
\newcommand{\eaz}{\end{eqnarray*}}
\newcommand{\bb}{\begin{thebibliography}}
\newcommand{\eb}{\end{thebibliography}}
\newcommand{\ct}[1]{${\cite{#1}}$}

\newcommand{\bi}[1]{\bibitem{#1}}

\begin{document}

\begin{center}
{\Large \bf  Instanton Contribution to   Polarized and
 Unpolarized
Gluon  Distributions in  Nucleon}
\footnote{  Talk presented at  DIS97 Workshop,
 April 1997, Chicago.}\\[1.5cm]
{ N.I.Kochelev}\\
{\it  DESY-Zeuthen
Platanenalle 6, D-15735 Zeuthen, Germany, and\\
JINR, Dubna, Moscow region, 141980,  Russia}\\

\end{center}

\begin{abstract}

The contribution of  the anomalous quark-gluon   interaction
induced by instantons  to the polarized and  unpolarized gluon
 distributions in nucleon is estimated.
  It is shown that this interaction  leads to
 {\it negative} gluon  polarization in nucleon.

\end{abstract}

\vskip 0.2cm
\section{Introduction}

In  recent years the interest in polarized hadron-hadron and
lepton-hadron
interactions at high energies has grown strongly.
 This interest stems from sensational result of measurement by the
 EMC(CERN) Collaboration \ct{EMC} in polarized DIS of the part of the
 proton spin carried by  quarks.
 It has been measured that this value is very small.
 As a result, a ``spin crisis`` of the naive parton
 picture   of the spin-dependent structure functions
 (see review \ct{rev}) arose.

 One of the  way to resolve this problem
 is  based on an assumption of the large  gluon polarization
in nucleon \ct{AA}.
 Recently,  NLO analysis
of the polarized DIS world data on $g_1(x,Q^2)$ was performed to
extract the polarized parton densities in  nucleon \ct{FortRud}.
The result suggests a {\it positive} value for the gluon polarization.
Still, this result is sensitive to the
input shapes of the polarized parton densities
as well as to the marginal statistical strength inherent in the $Q^2$
dependence of existing polarized DIS data.

A {\it positive} gluon polarization
in the proton is actually expected in the framework of
perturbative  QCD
due to conservation of helicity in  perturbative
quark-gluon vertex \ct{Brod}. However, in  QCD, the gluon distribution
function is a nonperturbative object and therefore one should
take into account the {\it nonperturbative} contribution to
  $\Delta G $.
Up to now only one calculation of the nonperturbative gluon
contribution to  $\Delta G$ was presented
in framework of  MIT bag model and the nonrelativistic model
\ct{jaffe}. It was shown that
the sign of the gluon polarization is {\it negative},
which is in  contradiction with the positive value
that one  would expect  to explain the ``spin crisis`` by a gluon
contribution only.

Here we  estimate the gluon polarization in  proton induced
by  nonperturbative vacuum fluctuations of the gluon fields,
so-called instantons \ct{Pol}.

\section{Anomalous Quark-Gluon Chromomagnetic Interaction Induced
by Instantons}

One of the models for the description of  non-perturbative
effects  in QCD is the instanton liquid model (see reviews \ct{a5},
\ct{a55}).
In this model  many properties of the hadrons as masses,
decays widths  etc., have been described rather well.

The existence of  instantons leads to a
specific spin--dependent t'Hooft's quark-quark interaction through
 QCD vacuum \ct{Hooft}, which
determines the spin-spin mass splitting in hadron multiplets
\ct{kochdor} and gives  rise to a negative sea quark polarization
and valence quark depolarization in  nucleon \ct{DorKoch}
(see also \ct{Forte}).

Recently, it was shown that instantons induce  the
 {\it anomalous chromomagnetic
quark-gluon} interaction \ct{koch1}:
\be
\Delta {\cal L_A}=
-i\mu_a
\sum_q\frac{g}{2m_q^*}\bar q\sigma_{\mu\nu}
t^a qG_{\mu\nu}^a.
\label{e4}
\ee
The value of the quark anomalous chromomagnetic moment can be
estimated in the liquid instanton model for  QCD vacuum \ct{a5} as
\be
\mu_a=-\frac{f\pi}{2\alpha_s},
\label{a6}
\ee
 where $f=n_c\pi^2\rho_c^4$ is the so-called packing fraction of instantons
 in  vacuum,
\\$m_q^*=m_q-2\pi^2\rho_c^2<0\mid \bar qq\mid 0>/3$ is
the effective quark mass,  and
$\rho_c $ is the average instanton size in the QCD vacuum.
The value of $n_c$ is connected with the value of the
gluon condensate by the formula:
\be
n_c=<0\mid \alpha_sG_{\mu\nu}^a G_{\mu\nu}^a\mid 0>/16\pi
\approx 7.5~10^{-4}{\  }GeV^4.
\nn
\ee

One can obtain  \re{e4} from the effective Lagrangian induced by instantons
 \ct{a6}
\begin{eqnarray} {\cal
L}_{eff}&=&\int\prod_q(m_q\rho-2\pi^2\rho^3\bar q_R(1+\frac{i}{4}
\tau^aU_{aa^\prime}\bar\eta_{a^\prime\mu\nu}\sigma_{\mu\nu})q_L)
\nonumber\\
&\cdot &exp^{-\frac{2\pi^2}{g}\rho^2U_{bb^{\prime}}
\bar\eta_{b^\prime\gamma\delta}
G^b_{\gamma\delta}}
\frac{d\rho}{\rho^5}d_0(\rho)d\hat{o}
+R\longleftrightarrow L,
\label{e1}
\end{eqnarray}
 where $\rho$ is the instanton size, $\tau^a$ are the matrices of the
 $SU(2)_c$ subgroup of the $SU(3)_c$ colour group,
 $d_0(\rho)$ is the density of the instantons, $d\hat{o}$ stands
 for integration over the instanton orientation in colour space,
$\int d\hat{o}=1$,
$U$ is the orientation matrix of the instanton,
 $\bar\eta_{a\mu\nu}$ is the numerical t'Hooft symbol and
 $\sigma_{\mu\nu}=[\gamma_\mu,\gamma_\nu]/2$.

 The  quark--gluon vertex \re{e4} follows from \re{e1} by expanding it
 in the set of  powers of the gluon strength and by integrating over
 $d\hat{o}$.

 For the numerical calculation, we use the NLO approximation for the strong
 coupling constant
\begin{equation}
\alpha_s(\rho)=-\frac{2\pi}{\beta_1t}(1+\frac{2\beta_2log t}{\beta_1^2t}) ,
\label{e17}
\end{equation}
 where
\begin{equation}
\beta_1=-\frac{33-2N_f}{6},\ \beta_2=-\frac{153-19N_f} {12}\nonumber
\end{equation}
 and
\begin{equation}
t=log(\frac{1}{\rho^2\Lambda^2}+\delta).
\label{e18}
\end{equation}
 In Equation (\ref{e18}), the parameter $\delta\approx1/\rho_c^2\Lambda^2$
 provides a smooth interpolation of the value of  $\alpha_s(\rho)$
 from the perturbative $(\rho\rightarrow0)$ to the nonperturbative
 region $(\rho\rightarrow\infty)$  \ct{a88}.

 For $N_f=3$, $\Lambda=230$MeV,
   the following estimate for
 the value of the anomalous quark chromomagnetic
moment has been obtained \ct{koch1}
\begin{equation}
\mu_a=-0.2 \ for \ \rho_c=1.6 {\rm GeV}^{-1}.
\nonumber
\end{equation}

\section{ Gluon Polarization Induced by Instantons}

The contribution of the interaction \re{e4} the gluon distribution
functions can be estimated by using the Altarelli-Parisi method \ct{AP}.
The matrix element for the transition of the initial
positive polarized quark state to  the final quark-qluon  state  has the
following form (see Fig.1):
\ba
|V_(q_+\rightarrow G_\pm q)|^2&=&\frac{C_F\mu^2_a}
{4{{m^*}_q}^2}
Tr(\hat{k_c}\sigma_{\mu\nu}\hat{k_a}\sigma_{\rho\tau}(1-\gamma_5))\nn\\
&\cdot &(q_\mu{\epsilon_\nu}^\pm-q_\nu{\epsilon_\mu}^\pm)^*
((q_\rho{\epsilon_\tau}^\pm-q_\tau{\epsilon_\rho}^\pm),
\label{matrix}
\ea
where ${\epsilon_\mu}^\pm$ is gluon polarization vector.

\begin{figure}
\centering
\epsfig{file=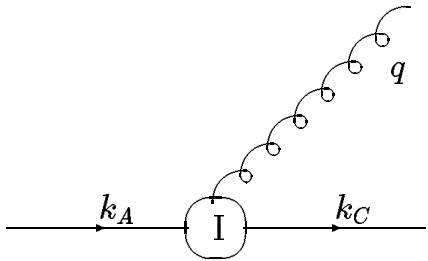,width=8.5cm}
\caption{\it The instanton contribution to the quark-gluon
splitting function.  }
\end{figure}

The calculation yields the result
\be
 |V_{q_+\rightarrow G_\lambda q}|^2=\frac{8p^4_\bot\mu_a^2}
{3{{m^*}_q}^2(z^2(1-z))}(1-\lambda),
\label{mat2}
\ee
where $z$ is the part of the initial quark momentum carried by the gluon,
$p_\bot$ is its transverse momentum~, $\lambda$ is the gluon
polarization. To obtain \re{mat2}, we have used the following
 quark and gluon momenta in the infinite momentum frame
 (see \ct{AP}):
\ba
k_A&=&(P,P,{\bf 0})\nn\\
k_C&=&((1-z)P+\frac{p^2_\bot}{2(1-z)P},(1-z)P,{\bf -p_\bot})\nn\\
q&=&(zP+\frac{p^2_\bot}{2zP},zP,{\bf p_\bot}).\nn
\ea
The result of the calculation of the quark  splitting function
averaged over the transverse momentum is

\be
P_{G_\lambda,q_+}(z,Q_0)=\frac{|\mu_a|(1-\lambda)}{8z}\int_0^{Q_0^2\rho_c^2}
d\beta, \label{split} \ee

where $\beta=p^2_\bot\rho_c^2 $,
and
 the relation  $2n_c=<0\mid\bar qq\mid 0>m^*_q {\ }$ \ct{a5} for the light
quarks ${\ }(m_q=0){\ }$ have been used.

It should be mentioned that due to  the spin-flip of
the quark at the instanton vertex the quark splitting function \re{split} has
a very specific dependence on the gluon helicity. It is non-zero only
if the emitted gluon has the opposite  helicity  compared to the
initial quark helicity.

 This  is in contrast to the
 case of the perturbative  quark-gluon vertex \ct{Brod}, where due to
 helicity conservation
the probability is larger to emit the gluon with the same helicity as the
initial quark.  As a result we anticipate a {\it positive}
gluon polarization induced by the perturbative quark-gluon vertex and
 {\it negative} gluon polarization induced by the non-perturbative
instanton-quark vertex.

To estimate the  instanton contribution to the polarized and
unpolarized gluon distribution,
the convolution formula  will be used
\be
\Delta G(x,Q_0)=\int_x^1\frac{dy}{y}\Delta P_{G,q}(y,Q_0)\Delta q_V
(\frac{x}{y}), \label{conv1} \ee

\be
 G(x,Q_0)=\int_x^1\frac{dy}{y} P_{G,q}(y,Q_0)q_V (\frac{x}{y}),
\label{conv2}
\ee
where
\be
\Delta P_{G,q}=P_{G_+,q_+}-P_{G_-,q_+},
 {\ }P_{G,q}=P_{G_+,q_+}+P_{G_-,q_+}.  \nn \ee For the unpolarized and
polarized valence quark distributions a simple  shapes were utilized:
 \ba
u_V(x)&=&2.18x^{-0.5}(1-x)^3, {\ }d_V(x)=1.09x^{-0.5}(1-x)^3\nn\\
\Delta u_V(x)&=&3.7(1-x)^3, {\ }\Delta d_V(x)=-1.3(1-x)^3.\label{val}
\ea
The unpolarized distributions have been normalized to the number of
$u-$ and $d-$ quarks in proton and the polarized ones have been normalized
 to the experimental data
on the weak decay coupling constants of  hyperons:

\be
g_A^3=\Delta u_V-\Delta d_V=1.25; {\ } g_A^8=\Delta u_V+\Delta d_V=0.6.
\label{coupl}
\ee

In the beginning we will estimate the instanton contribution at
 low value of
$Q_0\approx 1/\rho_c= 600 MeV$.

\begin{figure}[htb]
\centering
\epsfig{file=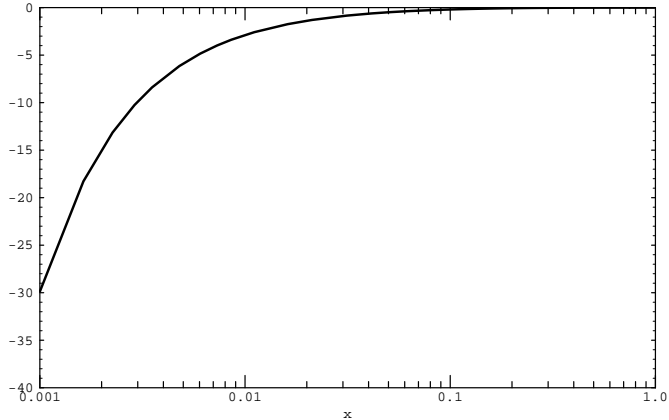,width=8.5cm}
\caption{\it The ~$x$  dependence of the gluon polarization
induced by instantons. }
\end{figure}

The results of the calculation of the polarized gluon distribution
is presented in Fig.~2. The most remarkable feature is that
the sign of the polarization  is negative  over all the kinematical range
and its size is steadily increasing
for decreasing $x$ values. The  total instanton contribution
to the gluon polarization in the nucleon  is
$\Delta G=-0.42$.

We can also estimate the instanton  contribution to the
{\it unpolarized} gluon distribution function in the same approach.
In this case due to the vanishing of the splitting function for the
same helicities of the initial quark and emitted gluon we have
\be
P_I(x,Q_0)=-\Delta P_I(x,Q_0),
\label{noni}
\ee
for the instanton contribution.

\begin{figure}[htb]
\centering
\psfig{file=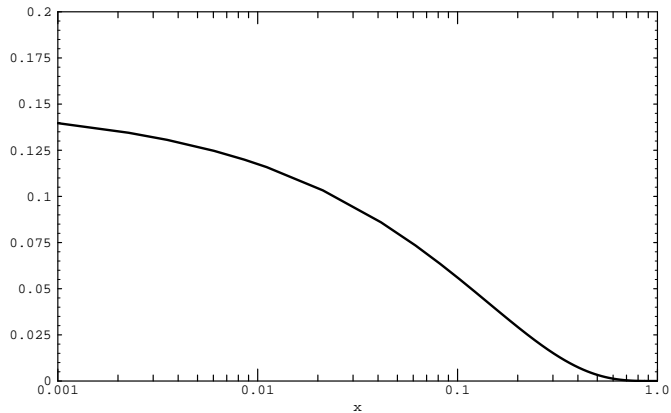,width=8.5cm}
\caption{\it The ~$x$  dependence of the unpolarized gluon distribution
$xG(x)$
induced by instantons. }
\end{figure}

The result of the calculation is presented in Fig.3.
The instanton contribution to the momentum of proton carried by gluons is
\be
\int_0^1dxxG_I(x,Q_0)=0.017.
\label{mom}
\ee

 One can take into account the $Q^2$ dependence of the instanton
contribution by using the simple formula
(see Eq.\re{split} in which $\mu_a\propto 1/\alpha_s $):
\be
\Delta G(x,Q^2)\approx \frac{\alpha_s(Q_0^2)}{\alpha_s(Q^2)}\Delta
G(x,Q_0),{\ } G(x,Q^2)\approx
\frac{\alpha_s(Q_0^2)}{\alpha_s(Q^2)}G(x,Q_0), \nn \ee with result at
$Q^2=4 GeV^2$ \be \Delta G_I(4 GeV^2)=-0.67, {\ } \int_0^1dxxG_I(x,4
GeV^2)=0.027.  \nn \ee

Therefore,  the instanton induced quark-gluon interaction
leads  to a rather large {\it negative} integral of the gluon
polarization
in nucleon. Furthermore, instanton induced glue gives about $5 \%$  to the
value of the proton momentum  carried by the gluons
\footnote{In Refs. \ct{insdis} and \ct{insdis2}
 some contribution from
 instantons to the coefficient functions of  unpolarized DIS has
been  taken into account in the diluty instanton gas approximation.}.

The negative gluon polarization
induced by instantons should gives, through axial anomaly \ct{AA},
 the {\it
positive } contribution to the spin-dependent structure functions
$g_1(x)$. Therefore we expect {\it positive} values for
neutron and proton structure functions $g_1^{p,n}(x)$ at low $x$
region where the instanton contribution  dominates.

As it was mentioned, recent NLO fits \ct{FortRud} of
the experimental data on $g_1^{p,n}(x,Q^2)$
show
the some indication on the {\it positive} value of
$\Delta G $. To explain the same experimental data with
the {\it negative} gluon polarization,  one
should introduce rather large {\it negative} sea quark polarization
 at  $x\geq 0.01$.
The fundamental mechanism for this polarization can be quark-quark
t'Hooft interaction induced by instantons (see \ct{DorKoch}).

\section{Summary}

In summary, we have shown that the instanton induced quark-gluon
interaction
leads  to a {\it large negative} gluon polarization in nucleon.
The instanton contribution to the  unpolarized gluon distribution
is  approximately $ 5\%$ at $Q^2=4GeV^2$.
The sign of a gluon polarization can be checked in HERMES and
COMPASS experiments by measurement of the double spin asymmetry in the
charm quarks production \ct{nowak}.

\section*{Acknowledgements}

The author is indebted to Prof. Paul S\"oding
for his permanent support of the present project. He also
thanks to J.Bl\"umlein, S.Brodsky, A.E.Dorokhov, B.L.Ioffe,
 T.Jegerlehner, T.Morii, M.Oka and A.Ringwald for helpful
discussions and DESY-Zeuthen for the hospitality extended to him.


\end{document}